\documentclass[aps,pra,twocolumn,floatfix,superscriptaddress]{revtex4}

\usepackage{amssymb,amsmath,amstext}                %%   American Physical Society math etc extensions
\usepackage{graphicx}                                               %%   Include figure files                                            %%   Manage links
\usepackage{epstopdf}                                               %%   help with eps -> pdf 
\usepackage{color}                                                     %%   change color
\usepackage{bm}                                                        %%   bold math                                    
\usepackage{appendix}                                              %%   formating appendicies
\usepackage[utf8]{inputenc}
\usepackage{bbold}
\usepackage{bbm}
\usepackage{ulem}
\usepackage{braket}
\usepackage{latexsym}
\usepackage[colorlinks=true,citecolor=blue,linkcolor=magenta]{hyperref}

\def\be{\begin{equation}}
\def\ee{\end{equation}}
\def\bea{\begin{eqnarray}}
\def\eea{\end{eqnarray}}

\def\bi{\begin{itemize}}
\def\ei{\end{itemize}}
\def\ben{\begin{enumerate}}
\def\een{\end{enumerate}}

\begin{document}

\title {Topologically protected quantized changes of the distance between atoms}

\author{Ali Emami Kopaei}
\thanks{ali.emami.app@gmail.com}
\affiliation{Szkoła Doktorska Nauk Ścisłych i Przyrodniczych, Wydział Fizyki, Astronomii i Informatyki Stosowanej, Uniwersytet Jagiello\'nski, ulica Profesora Stanisława Łojasiewicza 11, PL-30-348 Kraków, Poland}
\affiliation{Instytut Fizyki Teoretycznej, Uniwersytet Jagiello\'{n}ski, 
ulica Profesora Stanislawa Lojasiewicza 11, PL-30-348 Krak{ó}w, Poland}
\author{Krzysztof Giergiel}
\thanks{krzysztof.giergiel@gmail.com}
\affiliation{Instytut Fizyki Teoretycznej, Uniwersytet Jagiello\'{n}ski, 
ulica Profesora Stanislawa Lojasiewicza 11, PL-30-348 Krak{ó}w, Poland}
\affiliation{Optical Sciences Centre, Swinburne University of Technology, Melbourne
3122, Australia}
\affiliation{CSIRO, Manufacturing, Research Way, Clayton, Victoria 3168, Australia}
\author{Krzysztof Sacha}
\thanks{krzysztof.sacha@uj.edu.pl}
\affiliation{Instytut Fizyki Teoretycznej, Uniwersytet Jagiello\'{n}ski, 
ulica Profesora Stanislawa Lojasiewicza 11, PL-30-348 Krak{ó}w, Poland}
\affiliation{Centrum Marka Kaca, Uniwersytet Jagiello\'nski, ulica Profesora Stanisława Łojasiewicza 11, PL-30-348 Kraków, Poland}

\begin{abstract}
Thouless pumping enables the transport of particles in a one-dimensional periodic potential if the potential is slowly and periodically modulated in time. The change in the position of particles after each modulation period is quantized and depends solely on the topology of the pump cycle, making it robust against perturbations. Here, we demonstrate that Thouless pumping also allows for the realization of topologically protected quantized changes of the distance between atoms if the atomic s-wave scattering length is properly modulated in time.
\end{abstract}

\date{\today}

\maketitle

%\onecolumngrid

%\tableofcontents

\section{Introduction}
The study of time-dependent Hamiltonians has emerged as a vibrant and consequential field within the realm of quantum mechanics. These Hamiltonians play a crucial role in understanding the behavior of quantum systems governed by time-varying potentials, which can give rise to a diverse array of captivating and unforeseen phenomena. One particularly intriguing example that has garnered significant attention is the phenomenon known as Thouless pumping, first introduced by David J. Thouless in 1983 \cite{Thouless1983}.
Thouless pumping is a remarkable manifestation of time-dependent Hamiltonians, where the periodic modulation of the potential induces the transport of particles across a sample in a well-controlled and quantized manner. The transport is topologically protected and thus resistant to perturbations and has been realized experimentally \cite{Atala2013,Folling:2007p55101}.

On the other hand, when a periodically moving particle is resonantly driven by a periodically changing external force, an intriguing phenomenon arises. Its motion, when observed in a frame moving along an undisturbed periodic orbit, can be effectively described by a Hamiltonian reminiscent of solid-state physics. In this analogy, the effective Hamiltonian bears resemblance to that of an electron moving through a crystalline potential formed by ions \cite{Guo2013,Sacha15a}. What is interesting is that when this particle is observed in the laboratory frame, it exhibits solid-state behavior in the time domain \cite{Sacha15a,sacha16}. That is, changes in time of the probability for detecting the particle at a fixed position close to the periodic orbit reproduces the solid-state behavior described by the effective Hamiltonian in the moving frame. The potential in the effective Hamiltonian can be  engineered by skillfully selecting the Fourier components of the periodically changing external force. By doing so, it becomes possible to explore a wide range of condensed matter phenomena, which include Anderson localization, topological molecules, topological crystals, quasi-crystals, Mott-insulator phase, and many-body localization, all within the realm of time-domain observations \cite{Guo2013,Sacha15a,sacha16,Guo2016,Guo2016a,Giergiel2017,delande17,Mierzejewski2017,Liang2017,Giergiel2018,Giergiel2018a,Lustig2018,Giergiel2018b,Matus_2021,Giergiel2021,Giergiel2022,kopaei2022,kopaei2023}. This captivating domain of research is known as condensed matter physics in time crystals, with comprehensive reviews available \cite{Sacha2017rev,Guo2020,SachaTC2020,GuoBook2021}. It is worth emphasizing that time crystal investigations also encompass systems capable of spontaneously breaking time translation symmetry \cite{Wilczek2012,Bruno2013b,Watanabe2015,Syrwid2017,Kozin2019}. In the case of periodically driven systems, this implies the spontaneous breaking of discrete time translation symmetry dictated by the external drive, leading to the formation of discrete time crystals 
\cite{Sacha2015,Khemani16,ElseFTC,Giergiel2020,Wang2020,Wang2021,yarloo_2020}. Excitingly, experimental demonstrations of discrete time crystals have already been achieved \cite{Zhang2017,
Choi2017,Pal2018,Rovny2018,Smits2018,Mi2022,Randall2021,Frey2022,
Kessler2020,Kyprianidis2021,
Xu2021,Taheri2020,Bao2024,Kazuya2024,
Liu2024,Liu2024a}.

In this current research, we demonstrate a platform to observe the topological pumping of the particles: The modulation of the atomic s-wave scattering length in time. By adiabatically changing periodic modulation of the scattering length, we can realize quantized changes of the relative distance between atoms and this process is topologically protected. These findings contribute to the scientific understanding of controllable interactions and pave the way for further exploration of topological phenomena in experimental systems \cite{Walter2023,Arguello-Luengo2023}.

The paper is organized as follows. In Sec.~\ref{sec1} we introduce our system by considering two atoms moving on a one-dimensional (1D) ring and derive an effective Hamiltonian that describe resonante motion of the atoms.
In Sec.~\ref{sec2} we describe the Thouless pumping phenomenon that the effective Hamiltonian can reveal and explain what it means for the two-atom system.
Section~\ref{sec3} contains results of numerical simulations of the topologically protected control of the relative distance between the atoms.
The summary is given in Sec~\ref{conclusions}.
\section{Model
\label{sec1}}
\begin{figure}
    \centering
    \includegraphics[width=0.9\columnwidth]{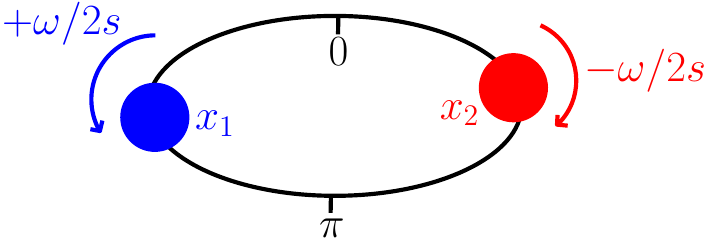}
    \caption{%Panel (a) 
    Presents schematic plot of the system considered in the paper. Two atoms are moving with the opposite velocity along the ring and are resonantly driven by periodic modulation of the s-wave scattering length.}
    \label{fig:my_label}
\end{figure}

Let us delve into the system of two atoms moving on a ring (see Fig.\ref{fig:my_label}). We shall assume that these atoms belong to the same species yet exhibit differences in their hyperfine states. When the kinetic energies are low, the interaction between these atoms can be effectively captured by means of a contact Dirac-delta potential with a strength, $2\pi g(t)$, proportional to the s-wave scattering length of the atoms. This interaction is periodically modulated in time which can be achieved through the utilization of either the Feshbach resonance \cite{Chin2010} or the confinement-induced resonance \cite{Olshanii1998}. Consequently, the Hamiltonian of the system takes the following form:
\begin{equation}
    H = \frac{p_1^2 + p_2^2}{2}  + 2\pi g(t) \delta(x_1 - x_2),
    \label{eq:1}
\end{equation}
where $x_i$ and $p_i$ represent the positions and conjugate momenta of the atoms on a ring, respectively. The periodic modulation $g(t)$ in time can read as follow,
\begin{equation}
g(t) = V_1 \sin^2(\omega t-\phi/2) + V_2 \cos^2(2\omega t).
\label{eq:2}
\end{equation}
The energy and length of the system are conveniently expressed in units of $\hbar^2/mR^2$ and $R$, respectively, where $R$ symbolizes the radius of the ring and $m$ represents the mass of the atoms. 

The interaction potential in Eq.(\ref{eq:1}) is dependent on the relative distance between the atoms, resulting in the decoupling of the center-of-mass degree of freedom from the relative coordinate. In this way, the  Hamiltonian can be expressed in terms of the center-of-mass coordinate, $X = (x_1 + x_2 )/2$, and the relative distance coordinate, $ x = x_1 - x_2 $, as well as their canonically conjugate momenta, $P = p_1 + p_2$ and $ p = (p_1 - p_2 )/2 $, respectively. The Hamiltonian of the system can be rewritten in terms of this coordinate as follows:
\begin{equation}
    H = H_{c.m.} + H_{rel},
    \label{eq:3}
\end{equation}
where
\begin{equation}
    H_{c.m.} = \frac{P^2}{4},
    \label{eq:4}
\end{equation}
\begin{equation}
    H_{rel} = p^2  + 2\pi g(t) \delta(x).
    \label{eq:5}
\end{equation}
According to Eq.(\ref{eq:4}), the center of mass degree of freedom exhibits behavior similar to the free particle. However, the relative coordinate exhibits more interesting behaviors especially when periodic modulation in $g(t)$ is resonant with the motion of the atoms along the ring. We assume the atoms are moving in the opposite velocities close to $\pm \omega/2s$ where $s$ is an integer number. To describe the system it is advantageous to consider the plane wave basis $\ket{n}$ as our chosen basis for the relative position degree of freedom and switch to a frame of reference that moves with the atoms using a unitary transformation, $H_{rel} \rightarrow U H_{rel}U^\dagger + i(\frac{\partial}{\partial t}U)U^\dagger$, where
\begin{equation}
U \ket{n} =
\begin{cases}
e^{-in\omega t/s}\ket{n}  & \text{for } n\ge 0, \\
e^{+in\omega t/s}\ket{n}  & \text{for } n<0.
\end{cases}
\label{eq:6}
\end{equation}
For the system in close proximity to the resonance, where the wave number $n$ is around $\pm \omega/2s$, we can obtain an effective time-independent Hamiltonian for the relative position degree of freedom by averaging the exact Hamiltonian over time (secular approximation) \cite{Berman1977,Lichtenberg1992,Buchleitner2002}. However, we have  to consider two resonance subspaces. In the first resonance subspace, where $n,n^{\prime}>0$, the time averaging leads to the following expression: 
\begin{eqnarray}
\begin{aligned}
		&H_{eff}^{(1,1)} =  [(n^2-n\omega/s) \delta_{n^{\prime},n}+ \Delta(n^{\prime},n)] \ket{n^{\prime}}\bra{n}
            \label{eq:6:1}
\end{aligned}
\end{eqnarray}
in which,
\begin{eqnarray}
		 \Delta(n^{\prime},n)&= 
           &V_1(\delta_{n^{\prime},n}/2 - \delta_{n^{\prime},n+2s} e^{-i\phi}/4-\delta_{n^{\prime},n-2s} e^{i\phi}/4)
		\cr &+&V_2(\delta_{n^{\prime},n}/2 + \delta_{n^{\prime},n+4s}/4 + \delta_{n^{\prime},n-4s}/4),
            \label{eq:6:1:1}
\end{eqnarray}
where $T=2s\pi/\omega$.
The same result will hold for considering the second subspace ($n,n^{\prime}<0$), $H_{eff}^{(2,2)} =  [(n^2+n\omega/s) \delta_{n^{\prime},n}+ \Delta(n^{\prime},n)] \ket{n^{\prime}}\bra{n}$. 
However, there will be a coupling between these two subspaces due to the zero-range Dirac-delta potential  \footnote{The coupling is not present if a sufficiently smooth version of the Dirac-delta potential is used \cite{Giergiel2018,SachaTC2020,Matus_2021}.}.
In the case of $n>0$ and $n^{\prime}<0$, 
\begin{eqnarray}
\begin{aligned}
		& H_{eff}^{(2,1)} = \Delta(-n^{\prime},n) \ket{n^{\prime}}\bra{n},
		\label{eq:7-1}
\end{aligned}
\end{eqnarray}
with $\Delta$ given in  Eq.(\ref{eq:6:1:1}), and similarly for  $n<0$ and $n^{\prime}>0$, ${H_{eff}^{(1,2)} = \Delta(n^{\prime},-n) \ket{n^{\prime}}\bra{n}}$.

Thus, the complete effective Hamiltonian can be expressed as,
\begin{equation}
\begin{aligned}
H_{eff} = \sum_{i,j=1}^{2} H_{eff}^{(i,j)}.
\label{eq:matrix}
\end{aligned}
\end{equation}
Apart from the omitted constant term, we can block diagonalize the Hamiltonian of Eq.(\ref{eq:matrix})
resulting in,
\begin{eqnarray}
{H}_{eff} &=&(m_1-\omega/2s)^2\; \delta_{m_1^{\prime},m_1}\ket{m_1^{\prime}}\bra{m_1} \label{eq:matrix1}\\
&+&
\left[(m_2-\omega/2s)^2\; \delta_{m_2^{\prime},m_2}+ 2\Delta(m_2^{\prime},m_2)\right] \ket{m_2^{\prime}}\bra{m_2},\nonumber
\end{eqnarray}
where $|m_{1,2}\rangle=(|n\rangle\mp|-n\rangle)/\sqrt{2}$. The constant terms $\omega/2s$ can be eliminated if we redefine $m_{1,2}\rightarrow m_{1,2}+\omega/2s$.
The first term of Eq.(\ref{eq:matrix1}), describes the free particle moving on the ring and the dynamics in the corresponding Hilbert subspace is trivial. 
However, the second part is non-trivial and the corresponding Hilbert subspace is a subspace where the Thouless pumping we are interested in can be realized. Looking at the form of the matrix elements of this part of the Hamiltonian it is not obvious that in the relevant Hilbert space one can realize the Thouless pumping. However, this part of the Hamiltonian is identical to the matrix elements of the following Hamiltonian calculated in the plane wave basis for $x$ in range of $(-2\pi,2\pi)$, 
\begin{equation}
{H}_{eff} =p^2 + 2 V_1 \sin^2\left(s x-\frac{\phi}{2}\right) + 2V_2 \cos^2\left(2sx\right),
\label{eq:8}
\end{equation}
and systems described by such a Hamiltonian can reveal the Thouless pumping (see the next section).

\begin{figure}
    \centering
    \includegraphics[width=1.0\columnwidth]{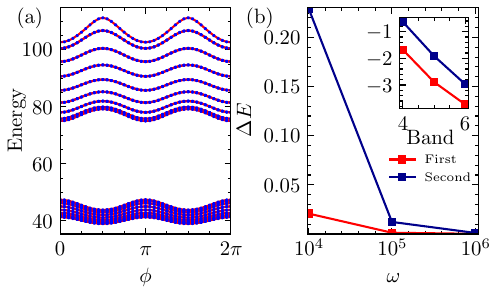}
    \caption{(a) Comparison between the quasi-energies of the exact Hamiltonian, Eq.~(\ref{eq:5}), (blue dots) and the energies of the effective Hamiltonian, Eq.~(\ref{eq:8}), (red lines) for $\omega = 10^5$, $V_1=32$, $V_2=16$, and $s=4$. The parameters were selected in order to make the energy levels structure clearly visible. 
 (b)  Differences, $\Delta E$, between the lowest eigenenergy levels for each band of the effective Hamiltonian and the corresponding exact quasi-energies of the system as a function of $\omega$ in a semi-log plot for $\phi=0$. Inset shows the same but in a log-log plot. If $\omega$ is sufficiently large, the agreement between the effective and exact results is perfect. The parameters used in this simulation are $V_1=64$, $V_2=32$, and $s=4$. }
    \label{fig:2}
\end{figure}

The spectrum of the system (\ref{eq:8}) as a function of the pumping parameter $\phi$ is demonstrated in Fig.~(\ref{fig:2})a. The spectrum contains two energy bands, and these bands are well separated from each other with a gap. Each band consists of $4s$ energy levels.
In Fig.(\ref{fig:2})b comparison of the spectra obtained by diagonalization of the exact Floquet Hamiltonian (\ref{eq:5})  and the effective Hamiltonian (\ref{eq:8}) is presented. If $\omega$ is sufficiently large, we obtain perfect agreement. This means that the phenomena that we can predict with the help of $H_{eff}$ are reproduced in the exact description of the system.

\section{Thouless Pumping
\label{sec2}}
The effective Hamiltonian, Eq.~(\ref{eq:8}), describes a system which can exhibit the Thouless pumping phenomenon which has been shown theoretically and demonstrated experimentally \cite{Atala2013,Peil:2003p61264,Folling:2007p55101,Trotzky:2008p464,viebahn2023interactioninduced,Greschner,Sergi2024,Ashirbad2023,Kuno_2020}.
Even though for some range of parameters, the system is well described by the tight-binding approximation, in the following we use the continuum model, Eq.~(\ref{eq:8}), \cite{Qian2011,Wang2013}.

The phenomenon of Thouless pumping is of significant interest as it offers a means of controlling the flow of particles in a system. This control is achieved by adiabatically  changing the control parameter, $\phi$ in Eq.~(\ref{eq:8}), which results in the transfer of quantized units of charge through the system. Here, it corresponds to changes of the average value $\langle x\rangle$ by the same amount after each adiabatic change of $\phi$ by $2\pi$.
The connection between a Chern topological invariant and charge pumping in the 1D system (\ref{eq:8}) is closely linked to the integer quantum Hall effect \cite{Thouless1982}. In the quantum Hall effect, charged particles move in a 2D space when a magnetic field is applied perpendicular to their motion. When the $\phi$ changes adiabatically, it is like threading magnetic flux through a cylinder. This creates an electric field in a perpendicular direction and results in quantized Hall conductance \cite{Thouless1983,Lohse2016}.
\begin{figure}
    \centering
    \includegraphics[width=0.9\columnwidth]{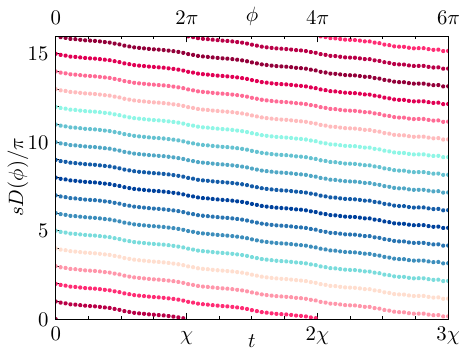}
    \caption{Illustration of the Thouless pumping described by the Hamiltonian (\ref{eq:8}) with the pumping control parameter $\phi$ changing adiabaticaly in time, i.e. $\phi$ changes by $2\pi$ over the pumping period $\chi=4\times 10^3T$. The figure shows quantized changes of the displacement, Eq.~(\ref{eq:10}). Different colors represent different Wannier states as the initial state. The parameters used in our simulation are $V_1=64$, $V_2=32$, and $s=4$.}
    \label{fig:3}
\end{figure}

As discussed in Ref.~\cite{Resta}, in the case of the ring geometry with the periodic boundary conditions, it is more convenient to analyze not the quantity $\braket{x}$ but 
\begin{equation}
D =2\mathrm{Im}\ln \bra{\psi}e^{ix/2}\ket{\psi},
\label{eq:9}
\end{equation}
which involves an operator explicitly respecting the periodic boundary conditions.
In this way, the position operator $D$ is represented in terms of a phase.
The introduction of a time-varying adiabatic parameter $\phi(t)$ yields the observable $\partial_t D(t)/2\pi = J(t)$ representing the particle current. Notably, the topological pumped charge \cite{Resta,Bardyn} $\Delta n$  after a cyclic protocol $t \in [0, \chi]$ with ${\phi(m\chi)-\phi(0) = m2\pi}$ is quantized as an integer and reads 
\begin{equation}
\Delta n =  \frac{s}{\pi} \left[D(\phi=2m\pi) - D(\phi=0)\right].
\label{eq:10}
\end{equation}  
Figure~\ref{fig:3} illustrates the Thouless pumping. It shows how $D$ evolves in time when different Wannier states of the first energy band of the Hamiltonian (\ref{eq:8}) are chosen as initial states and $\phi(t)$ changes adiabatically in time, i.e., $\phi(t)$ changes by $2\pi$ over the period $\chi=4\times10^3T$ which turns out to be sufficiently longer than the minimal tunneling time $2\times10^3 T$ between neighboring lattice sites of the Hamiltonian (\ref{eq:8}).

Typically a Hamiltonian of the form of Eq.~(\ref{eq:8}) is related to particles in an external spatially periodic potential which can reveal quantized translation of the center of mass position of the particles when the control parameter $\phi$ is adiabatically changed by $2\pi$. Here, the meaning of such a Thouless pumping is different because $x$ in Eq.~(\ref{eq:8}) describes the relative distance between two atoms. In the following we show that the Thouless pumping allows us to control quantized changes of the relative distance between the atoms which are topologically protected.

\begin{figure}
    \includegraphics{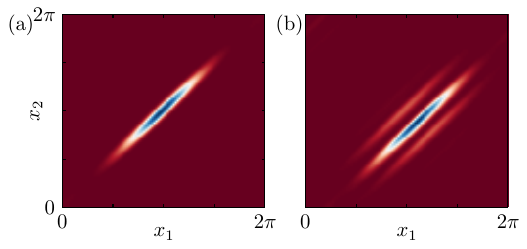}
    \caption{Time evolution of the two-atom system. The initial wavefunction has been chosen as a product of a Gaussian function for the center of mass degree of freedom (centered at $X=\pi$ and with the standard deviation $\sigma_{c.m.}=0.75$) and a single Wannier state of the first energy band of the Hamiltonian (\ref{eq:8}) (with $s=4$ and $\phi=0$) for the relative degree of freedom.  
     Panel (a) shows the probability density in the laboratory frame at $t=0$. Panel (b) shows the same but at $t=\chi=4\times10^3T$, i.e. after one pumping cycle when $\phi$ changes from 0 to $2\pi$. The maximum of the probability density along the direction of the relative position of the atoms (diagonal of the plot) is shifted by $\pi/4$ indicating the quantized Thouless pumping. The parameters are the same as in Fig.(\ref{fig:3}).}
    \label{fig:4}
\end{figure}

\section{adiabatic evolution
\label{sec3}}
In this section we present results of the evolution of two atoms on the ring with periodic modulation of the contact interaction strength, $2\pi g(t)$, and an additional adiabatic changes of the pumping control parameter $\phi$, see Eq.~(\ref{eq:2}) and Eq.~(\ref{eq:8}). 

\begin{figure}
    \centering
    \includegraphics[width=0.9\columnwidth]{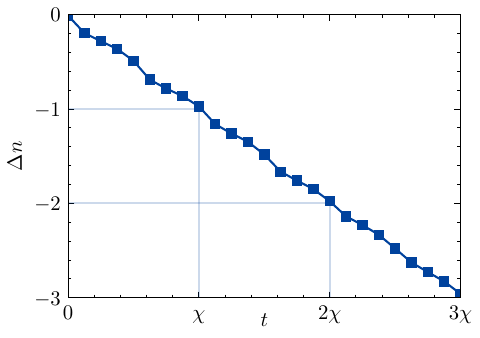}
    \caption{Similar results as presented in Fig.~\ref{fig:4} but for the initial state of the relative degree of freedom of the atoms being a superposition of the Wannier states with a Gaussian envelop centered at $x=0$ and with the standard deviation $\sigma_{rel}=0.17$. After each pumping cycle (i.e., at $t=\chi$, $2\chi$ and $3\chi$ in the plots), the probability density along the direction of the relative position of the atoms shifts by the same value indicating quantization of $\Delta n$, Eq.~(\ref{eq:10}).  The parameters are the same as in Fig.~\ref{fig:3}.}
    \label{fig:5}
\end{figure}

The center of mass degree of freedom of two atoms, denoted by $X$, is described by the free-particle like Hamiltonian (\ref{eq:4}) while the relative position degree of freedom of the atoms by the Hamiltonian (\ref{eq:5}) which in the moving frame and in the relevant Hilbert subspace can be approximated by the effective time-independent Hamiltonian (\ref{eq:8}).
%\aek{ Starting from a single point $(X,x)$ it takes $4\pi$ distance moving in the $x$ direction to return to the stating point. The corresponding first energy band of (\ref{eq:8}) consists of $4s$ states. They form $2s$ pairs of states for $x$ in the range $[0,2\pi)$ connected by the change of subspace of $X$ degree of freedom.} \ks{[KS: Do we realy need these sentences?]}
We assume that the initial state is a product, $\psi_{c.m.}(X)\psi_{rel}(x)$, of a Gaussian wavefunction for the center of mass degree of freedom, $\psi_{c.m.}(X)\propto \exp[-(X-\pi)^2/2\sigma_{c.m.}^2]$ with $\sigma_{c.m.}=0.75$, and a single Wannier state, $\psi_{rel}(x)$, of the first energy band of the Hamiltonian (\ref{eq:8}) with $\phi=0$ for the relative degree of freedom. Results of the time evolution in the laboratory frame with the adiabatically changing pumping parameter $\phi(t)$ [i.e., $\phi(t)$ changes by $2\pi$ over the time period $\chi=4\times10^3T$ are presented in Fig.~\ref{fig:4}. While the probability density spreads along the direction of the center of mass position (anti-diagonal direction in Fig.~\ref{fig:4}), along the direction of the relative position of the atoms we can see that the maximum of the probability density is shifted by $\pi/4$. The latter means that the average distance between the atoms has changed from $0$ to $\pi/4$. Each next pumping cycle increases the average relative distance always by $\pi/4$ and such quantized manipulation of the relative particles' distance is topologically protected. Results of three consecutive pumping cycles are presented in Fig.~\ref{fig:5} but this time the initial wavefunction for the relative degree of freedom, $\psi_{rel}(x)$, has been chosen as a Gaussian superposition of the neighboring Wannier states centered at $x=0$ and with the standard deviation $\sigma_{rel}=0.17$. Similarly like in Fig.~\ref{fig:4} the probability distribution is spreading due to the tunneling between lattice sites of the potential in (\ref{eq:8}) but the average displacement after each pumping cycle is always $\pi/4$.

The quantized changes of the distance between atoms can be performed in ultra-cold atoms. To give a taste of the experimental parameters, let us consider the following example. Assume that two $^{39}$K atoms in the two different hyperfine states, $\ket{F=1,m_F=0}$ and $\ket{F=1,m_F=-1}$, are trapped in the quasi-1D ring with the toroidal trap of the radius $R=40~\mu$m and the transverse harmonic confinement of the frequency $\omega_\perp=2\pi \times 10$~kHz,
 corresponding to the transverse radius 15~nm \cite{Wright2000}. The atoms are initially prepared in the Gaussian wavepackets of the width $\sigma=6.8~\mu$m located on the opposite sides of the ring and with the average velocities $\pm 25.1$~mm/s. The interspecies s-wave scattering length can be modulated in time by changing the magnetic field close to the Feshbach resonance at $113.76$~G \cite{Tanzi2018,DErrico_2007} or by means of confinement induced resonances \cite{Chin2004}. If the scattering length is periodically modulated with frequency $\omega=3.2\pi$~kHz and amplitude $7.9$~nm (which corresponds to $V_2=4.0$) 
 and the modulation consists of $k_m=8$ harmonics for $s=2$, 
 then the quantized changes of the relative distance between the atoms, we consider here, can be observed. In order to realize the effective Hamiltonian, one has to apply time periodic driving with frequencies $\omega$ and $2\omega$. This requirement can be easily fulfilled in the experiment because having the driving with the frequency $\omega$ its second harmonic with the frequency $2\omega$ can be generated.
\section{Summary\label{conclusions}}
We consider the behavior of a two-atom system moving on a ring where the interaction between the atoms is described using the contact Dirac-delta potential and its strength is modulated in time by means of either Feshbach resonance or the confinement induced resonance. 
When the atoms are moving at opposite velocities which fulfill the resonance condition, i.e., when the driving frequency is an integer multiple of the frequency of the motion of the atoms, we can apply the secular approximation. Then, in the frame moving with the atoms, the system’s behavior can be accurately reproduced by a time-independent effective Hamiltonian.
Notably, the potential in the effective Hamiltonian can be engineered to create the Thouless pumping phenomenon by properly selecting the Fourier components of the periodically changing s-wave scattering length of the atoms. This effective potential emerges from resonant connections between various harmonics of atoms translational motion and the time-periodic modulation of the original atom-atom interactions.
We demonstrated that if we initialize the relative position degree of freedom of the atoms in a localized Wannier state or superposition of the Wannier states with a Gaussian envelope, the system exibits the topologically protected quantized changes of the distance between atoms. 
It is important to emphasize that, in our system, the Thouless pumping is distinct from the usual situation. Here, the quantized changes are intricately connected to the relative distance between the two atoms, instead of the center of mass position of particles, as the pumping control parameter undergoes an adiabatic change. 

\section{ACKNOWLEDGMENTS}
We acknowledge the support of the
National Science Centre, Poland, via Project No. 2021/42/A/ST2/00017. The research has been supported by a grant from the Priority Research Area (DigiWorld) under the Strategic Programme Excellence Initiative at Jagiellonian University. The numerical computations in this work were supported in part by PL-Grid Infrastructure, Project No. PLG/2023/016644.
Some parts of our numerical schemes are based on the PETSc \cite{petsc-user-ref,petsc-efficient} and SLEPc \cite{Hernandez:2005} libraries.

\bibliography{ref_tc_book}
\end{document}